\documentclass[aps,prc,twocolumn,showpacs,amsmath,amssymb,superscriptaddress]
{revtex4}

\usepackage{graphicx}
\usepackage{dcolumn}
\usepackage{bm}
\usepackage{enumerate}
\usepackage{amsmath}
\usepackage{amssymb}
\usepackage{tabularx}
\usepackage{mathtools}
\usepackage{xcolor}
\usepackage{ctable}
\usepackage{rotating}
\usepackage{wrapfig}

\begin{document}

\preprint{APS/PRC - Nuclear Reaction}

\title{Search for stabilizing effects of $\bm{Z=82}$ shell closure against
fission}

\author{J. Gehlot}
\affiliation{Nuclear Physics Group, Inter University Accelerator Centre,
Aruna Asaf Ali Marg, Post Box 10502, New Delhi 110067, India}
\author{S. Nath}
\email{subir@iuac.res.in}
\affiliation{Nuclear Physics Group, Inter University Accelerator Centre,
Aruna Asaf Ali Marg, Post Box 10502, New Delhi 110067, India}
\author{Tathagata Banerjee}
\altaffiliation{Presently at Department of Nuclear Physics, Research School of
Physics and Engineering, The Australian National University, Canberra ACT 2601,
Australia.}
\affiliation{Nuclear Physics Group, Inter University Accelerator Centre,
Aruna Asaf Ali Marg, Post Box 10502, New Delhi 110067, India}
\author{Ish Mukul}
\altaffiliation{Presently at TRIUMF, 4004 Wesbrook Mall, Vancouver,
British Columbia, V6T 2A3, Canada.}
\affiliation{Nuclear Physics Group, Inter University Accelerator Centre,
Aruna Asaf Ali Marg, Post Box 10502, New Delhi 110067, India}
\author{R. Dubey}
\altaffiliation{Presently at iThemba LABS, National Research Foundation,
PO Box 722, 7129 Somerset West, South Africa.}
\affiliation{Nuclear Physics Group, Inter University Accelerator Centre,
Aruna Asaf Ali Marg, Post Box 10502, New Delhi 110067, India}
\author{A. Shamlath}
\affiliation{Department of Physics, School of Mathematical and Physical
Sciences, Central University of Kerala, Kasaragod 671314, India}
\author{P. V. Laveen}
\affiliation{Department of Physics, School of Mathematical and Physical
Sciences, Central University of Kerala, Kasaragod 671314, India}
\author{M. Shareef}
\affiliation{Department of Physics, School of Mathematical and Physical
Sciences, Central University of Kerala, Kasaragod 671314, India}
\author{Md. Moin Shaikh}
\altaffiliation{Presently at Variable Energy Cyclotron Centre, 1/AF Bidhan
Nagar, Kolkata 700064, India.}
\affiliation{Nuclear Physics Group, Inter University Accelerator Centre,
Aruna Asaf Ali Marg, Post Box 10502, New Delhi 110067, India}
\author{A. Jhingan}
\affiliation{Nuclear Physics Group, Inter University Accelerator Centre,
Aruna Asaf Ali Marg, Post Box 10502, New Delhi 110067, India}
\author{N. Madhavan}
\affiliation{Nuclear Physics Group, Inter University Accelerator Centre,
Aruna Asaf Ali Marg, Post Box 10502, New Delhi 110067, India}
\author{Tapan Rajbongshi}
\altaffiliation{Presently at Central Research Hub, Assam Science and Technology
University, Guwahati 781013, India.}
\affiliation{Department of Physics, Gauhati University, Guwahati 781014, India}
\author{P. Jisha}
\affiliation{Department of Physics, University of Calicut, Calicut 673635,
India}
\author{Santanu Pal}
\altaffiliation{Formerly with Physics Group, Variable Energy Cyclotron Centre,
1/AF Bidhan Nagar, Kolkata 700064, India.}
\affiliation{Nuclear Physics Group, Inter University Accelerator Centre,
Aruna Asaf Ali Marg, Post Box 10502, New Delhi 110067, India}

\date{\today}

\begin{abstract}
\begin{description}
\item[Background:]
Presence of closed proton and/or neutron shells causes deviation from
macrocopic properties of nuclei which are understood in terms of the liquid
drop model. Efforts to synthesize artificial elements are driven by prediction
of existence of closed shells beyond the heaviest doubly magic nucleus found in
nature. It is important to investigate experimentally the stabilizing effects
of shell closure, if any, against fission.
\item[Purpose:] This work aims to investigate probable effects of proton shell
($Z = 82$) closure in the compound nucleus, in enhancing survival probability
of the evaporation residues formed in heavy ion-induced fusion-fission
reactions.
\item[Method:] Evaporation residue cross sections have been measured for the
reactions $^{19}$F+$^{180}$Hf, $^{19}$F+$^{181}$Ta and $^{19}$F+$^{182}$W from
$\simeq9\%$ below to $\simeq42\%$ above the Coulomb barrier; leading to
formation of compound nuclei with same number of neutrons ($N = 118$) but
different number of protons across $Z = 82$; employing the Heavy Ion Reaction
Analyzer at IUAC. Measured excitation functions have been compared with
statistical model calculation, in which reduced dissipation coefficient is the
only adjustable parameter.
\item[Results:] Evaporation residue cross section, normalized by capture cross
section, is found to decrease gradually with increasing fissility of the
compound nucleus. Measured evaporation residue cross sections require inclusion
of nuclear viscosity in the model calculations. Reduced dissipation coefficient
in the range of 1\textendash3 $\times$ $10^{21}$ s$^{-1}$ reproduces the data
quite well.
\item[Conclusions:] Since entrance channel properties of the reactions and
structural properties of the heavier reaction partners are very similar, degree
of presence of non-compound nuclear fission, if any, is not expected to be
significantly different in the three cases. No abrupt enhancement of
evaporation residue cross sections has been observed in the reaction forming
compound nucleus with $Z = 82$. Thus, this work does not find enhanced
stabilizing effects of $Z = 82$ shell closure against fission in the compound
nucleus. One may attempt to measure cross sections of individual exit channels
for further confirmation of our observation.
\end{description}
\end{abstract}


\maketitle

\section{Introduction}
\label{intro}
Bohr and Wheeler \cite{Bohr1939} modelled the atomic nucleus as a homogeneously
charged liquid drop. Many macroscopic properties of nuclei, most notably the
phenomenon of fission \cite{Fission1939}, in which a heavy nucleus splits
itself into lighter fragments, could be understood in terms of the liquid drop
model. However, limitations of this model to explain microscopic features,
\textit{e.g.} enhanced stability of a few nuclei, led to development of the
nuclear shell model by Mayers and others \cite{Shell19484950}. Since then,
effects of shells on nuclear reaction dynamics has been a topic of great
interest. Most significantly, superheavy nuclei, beyond the heaviest nucleus
available in nature, have been hypothesised to exist solely because of shell
stabilization effects. Sustained efforts in the field of heavy element
research, since the first prediction \cite{Sobiczewski1966} of a doubly
shell-closed nucleus beyond $^{208}_{82}$Pb$_{126}$, culminated recently into
completion of the seventh period of the periodic table of elements
\cite{iupac-ptable}. Though the trans-lead doubly shell-closed nucleus is yet
to be synthesized in a laboratory, the cardinal role of shell stabilization in
enhancing life time of superheavy nuclei has been firmly established
\cite{Hamilton2013}.

\begin{table*}[ht!]
\label{params}
\begin{center}
\caption{Details of the nuclear reactions studied in this work. $\beta_{2}$,
$V_{\textrm{B}}$, $Q_{\textrm{CN}}$, $\chi_{\textrm{CN}}$ and
$\eta_{\textrm{BG}}$ are the quadrupole deformation, the Coulomb barrier,
$Q$-value of the reaction, CN fissility and the Businaro-Gallone critical mass
asymmetry, respectively.}
\begin{tabular}{l l l l l l l l l l l}
\hline\hline
System & $\beta_{2}$ (target) & $V_{\textrm{B}}$ & $Z_{\textrm{p}}Z_{\textrm{t}}$ & $\eta$ & CN & $Q_{\textrm{CN}}$ & $\chi_{\textrm{CN}}$ & $\eta$$_{\textrm{BG}}$ \\ [0.5ex]
 & & (MeV) & & & & (MeV) & & \\ [0.5ex]
\hline
$^{19}_{9}$F$_{10}$+$^{180}_{72}$Hf$_{108}$ & 0.274 & 76.8 & 648 & 0.809 & $^{199}_{81}$Tl$_{118}$ & -23.210 & 0.691 & 0.831 \\
$^{19}_{9}$F$_{10}$+$^{181}_{73}$Ta$_{108}$ & 0.269 & 77.9 & 657 & 0.810 & $^{200}_{82}$Pb$_{118}$ & -23.678 & 0.701 & 0.838 \\
$^{19}_{9}$F$_{10}$+$^{182}_{74}$W$_{108}$ & 0.259 & 79.0 & 666 & 0.811 & $^{201}_{83}$Bi$_{118}$  & -28.314 & 0.712 & 0.844 \\[1ex]
\hline
\end{tabular}
\end{center}
\end{table*}

Formation cross sections of superheavy evaporation residues (ERs) being
vanishingly small, it is rather challenging to study the dynamics of such
reactions. Several studies on effects of shell closure on reaction dynamics,
therefore, have been reported in the mass region around
$^{208}_{82}$Pb$_{126}$. One important difference
between the nuclei in the vicinity of $Z = 82$, $N = 126$ and the superheavy
nuclei, though, should be borne in mind. While the fission barrier in the
latter arises solely because of shell effects, the liquid drop model accounts
for a substantive part of the fission barrier in the former. The first
comprehensive investigation to verify reduction of fission competetion in
deexcitation of the compound nucleus (CN) due to stabilizing influence of the
strong ground-state shell effect in the vicinity of $N = 126$ was reported by
Vermeulen \textit{et al.} \cite{DVermeulen1984}. However, the results showed
`surprisingly' low stabilizing influence of the spherical shell against fission
competetion. Andreyev \textit{et al.} \cite{ANAndreyev2005} studied systematics
of ER cross sections ($\sigma_{\textrm{ER}}$) for the neutron-deficient CN
$^{184 \textendash 192}_{83}$Bi$^{*}$ and $^{186 \textendash 192}_{84}$Po$^{*}$
formed in complete fusion between two heavy ions. A satisfactory reproduction
of the data by the statistical model demanded up to 35\% reduction of the
fission barrier. Based on the systematic analysis, the authors concluded
`strongly' increased fissility above the shell closure at $Z = 82$. Nath
\textit{et al.} measured $\sigma_{\textrm{ER}}$ \cite{SNath2010} and ER-gated
CN angular momentum ($\ell$) distribution \cite{SNath2011} for
$^{19}$F+$^{184}$W. The results were further compared with those from
neighbourng systems with nearly similar entrance channel charge product,
$Z_{\textrm{p}}Z_{\textrm{t}}$ and mass asymmetry,
$\eta = \frac{|A_{\textrm{p}}-A_{\textrm{t}}|}{A_{\textrm{p}}+A_{\textrm{t}}}$
(here $Z_{\textrm{p}}$ ($Z_{\textrm{t}}$) and $A_{\textrm{p}}$
($A_{\textrm{t}}$) denote atomic number and mass number for the projectile
(target), respectively). The fission barrier for the CN with $Z = 82$ was found
to deviate from the systematic ($N$,$Z$) dependence. Similar measurements were
carried out for the reactions $^{30}$Si+$^{170}$Er and $^{31}$P+$^{170}$Er
forming the CN $^{200}_{82}$Pb$^{*}_{118}$ and $^{201}_{83}$Bi$^{*}_{118}$,
respectively, by Mohanto \textit{et al.} \cite{GMohanto2012,GMohanto2013}. The
results revealed no clear signature of extra stability due to $Z = 82$ shell
closure, showing similar $\sigma_{\textrm{ER}}$ and moments of
$\ell$-distribution for both the reactions at a given
$E_{\textrm{c.m.}} - V_{\textrm{B}}$.

These works relied upon statistical model of decay of CN to interpret the data.
This approach is questionable, in some cases, as reactions induced by heavier
projectiles (\textit{e.g.} $^{40}$Ar \cite{DVermeulen1984} and $^{46}$Ti,
$^{50,52}$Cr, $^{94,95,98}$Mo \cite{ANAndreyev2005}) have been known to go
through non-equilibrium processes like quasi-fission, thereby inhibiting
formation of CN, equilibrated in all degrees of freedom. There are many recent
studies in support of this argument \cite{RduRietz2013,AShamlath2017}.
The statistical models used by various groups of researchers also differ in
details. To explain absence of `expected' stabilization against fission for
spherical nuclei near $N = 126$, Junghans \textit{et al.} \cite{ARJunghans1998}
included collective enhancement of level density (CELD) in the calculation.
\textit{Ad hoc} reduction of fission barrier was also suggested to reproduce
measured $\sigma_{\textrm{ER}}$ \cite{ANAndreyev2005,GMohanto2013}.

In this article, we revisit the question whether $Z = 82$ shell closure
enhances survival of ERs against fission. To improve upon earlier attempts,
we have chosen three reactions to form CN with same number of neutrons
($N = 118$) and different numbers for protons across $Z = 82$ (see Table
\ref{params}). The facts \textendash (a) the reactions are induced by $^{19}$F
projectiles and (b) entrance channel parameters of the three reactions are
nearly the same \textendash lower the possibility of non-CN fission (NCNF)
affecting ER formation significantly and with varying degree of severity in the
three reactions. It is well known that shell effects tend to disappear at
higher excitation energy ($E^{*}$). Recent measurements of fission fragment (FF)
mass distribution from heavy CN \cite{AChaudhuri2015sh,KNishio2015} points to a
\textit{threshold} of $E^{*} \sim 40$ MeV, up to which shell effects persist.
The three CN are formed with $E^{*}$ in the range of 42\textendash92 MeV in the
present experiment. The statistical model calculations performed in this work
include all important physical phenomena known to affect fission dynamics and
have a single adjustable parameter, \textit{viz.} reduced dissipation
coefficient, $\beta$. Thus, scrutiny of results of the three reactions is
expected to bring forth stabilizing influences of $Z = 82$ shell closure
against fission, if any.

The article is organized as follows. The experiment is described in Sec.
\ref{Expt}. Results from the experiment and model calculations are presented in
Sec. \ref{Results}. Section \ref{Discuss} contains a discussion followed by
summary and conclusion in Sec. \ref{Conc}.

\section{The Experiment}
\label{Expt}
The experiment has been carried out at the 15UD Pelletron accelerator facility
of IUAC, New Delhi. A pulsed $^{19}$F beam, with pulse separation of 4 $\mu$s,
has been incident upon $^{180}$Hf (150 $\mu$g/cm$^2$), $^{181}$Ta
(175 $\mu$g/cm$^2$) and $^{182}$W (70 $\mu$g/cm$^2$) targets, all with thin
($\sim$20 $\mu$g/cm$^2$) $^{\textrm{nat}}$C backing \cite{TBVac2018}. Important
parameters of the three reactions are listed in Table \ref{params}. ER cross
sections ($\sigma_{\textrm{ER}}$) have been measured, employing the recoil mass
spectrometer Heavy Ion Reaction Analyzer (HIRA) \cite{HIRA}, at projectile
energies ($E_{\textrm{lab}}$) in the range of 80\textendash124 MeV. Two silicon
detectos, placed inside the target chamber at
$\theta_{\textrm{lab}} = 15.5^{\circ}$ with resect to the beam direction and in
the horizontal plane, have been used for monitoring position of the beam on
targets and absolute normalization of $\sigma_{\textrm{ER}}$. ERs have been
separated from the overwhelmingly-dominant background events by the HIRA and
transported to its focal plane. To ensure that charge states of ERs follow the
equilibrium distribution, a thin ($\sim30$ $\mu$g/cm$^2$) $^{\textrm{nat}}$C
foil has been placed at $\theta_{\textrm{lab}} = 0^{\circ}$, 10.0 cm downstream
from the target. The ERs have been detected by a multi-wire proportional
counter (MWPC), placed at the focal plane of the spectrometer. The MWPC, with
an active area of 15.0$\times$5.0 cm$^{2}$ and a mylar window of thickness 0.5
$\mu$m, has been operated with isobutane at 3 mbar pressure. Measuremets have
been performed keeping the HIRA at $\theta_{\textrm{lab}} = 0^{\circ}$ and with
full acceptance of 10 msr. Time of flight (TOF) of the ERs, over the distance
from the target to the MWPC, have also been recorded. List mode data have been
collected with the logical OR of the timing signals from the MWPC and the two
monitor detectors as the master trigger.

\begin{figure}[ht!]
\includegraphics[scale=0.425]{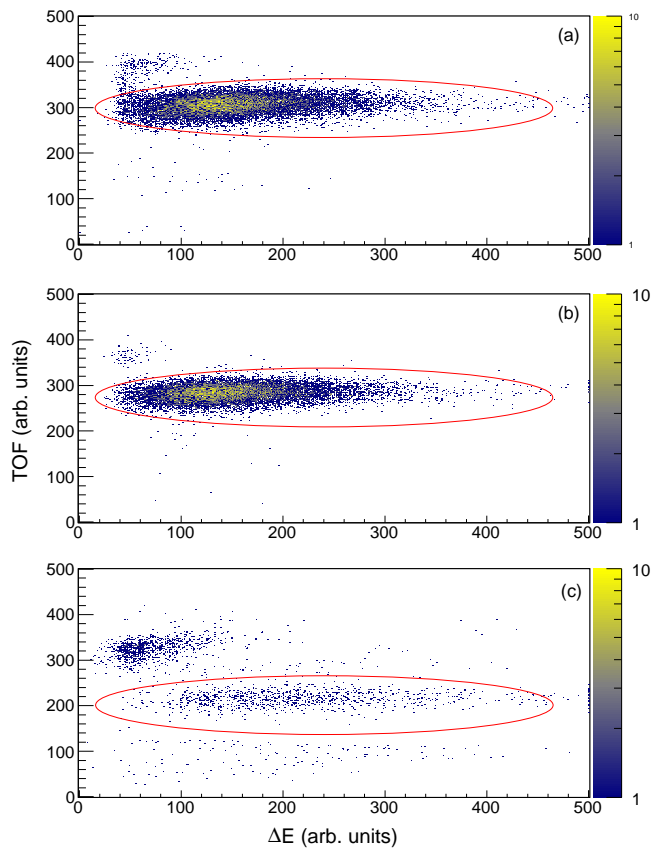}
\caption{\label{ScattPlots} Scatter plots between $\Delta E$ and TOF of the events
recorded at the focal plane of the HIRA for
(a) $^{19}$F + $^{180}$Hf at $E_{\textrm{lab}} = 119.7$ MeV
$\left(\frac{E_{\textrm{c.m.}}}{V_{\textrm{B}}} \simeq 1.41\right)$,
(b) $^{19}$F + $^{181}$Ta at $E_{\textrm{lab}} = 99.6$ MeV
$\left(\frac{E_{\textrm{c.m.}}}{V_{\textrm{B}}} \simeq 1.16\right)$ and
(c) $^{19}$F + $^{182}$W at $E_{\textrm{lab}} = 79.6$ MeV
$\left(\frac{E_{\textrm{c.m.}}}{V_{\textrm{B}}} \simeq 0.91\right)$.
$E_{\textrm{c.m.}}$ stands for energy available in the centre of mass (c.m.)
frame of reference. ER events are enclosed within an elliptical gate in each
plot.}
\end{figure}

\section{Data Analysis and Results}
\label{Results}
\subsection{ER cross sections}
The first step towards experimental determination of $\sigma_{\textrm{ER}}$ is
to identify the ERs unambiguously at the focal plane of the spectrometer. This
is achieved by generating scatter plots between energy loss of ERs ($\Delta E$)
at the focal plane and the TOF signals. Three such plots for the three
reactions are shown in Fig. \ref{ScattPlots}. Inherent backgroud rejection
capability of the HIRA, for very asymmetric reactions like the present ones,
ensures that the ERs can be clearly separated from the few projectile-like
particles reaching the focal plane. It is generally observed that the intensity
of background events at the focal plane of the HIRA, though insignificant in
most cases, increases gradually with decrease in projectile energy. However,
quite satisfactory separation between ERs and background events has been
obtained over the entire range of $E_{\textrm{lab}}$ in the present experiment,
as is evidenced by the $\Delta E$ \textendash TOF plot for $^{19}$F+$^{182}$W
at the lowest $E_{\textrm{lab}}$, shown in panel (c) of Fig. \ref{ScattPlots}.

The second most important aspect in the analysis is to estimate efficiency of
HIRA $\epsilon_{\textrm{HIRA}}$. Only a fraction of ERs, produced in a fusion
reaction, reaches the focal plane and is recorded by the detector.
$\epsilon_{\textrm{HIRA}}$ for the ERs varies depending upon several reaction
parameters. The same has been calculated employing the semi-microscopic Monte
Carlo code \textsc{ters} \cite{SNath20082009} following the formalism outlined
in Ref. \cite{SNath2010}.

\begin{figure*}[ht!]
\includegraphics[width=0.95\textwidth,trim=0.0cm 0.0cm 0.0cm 0.0cm,clip]{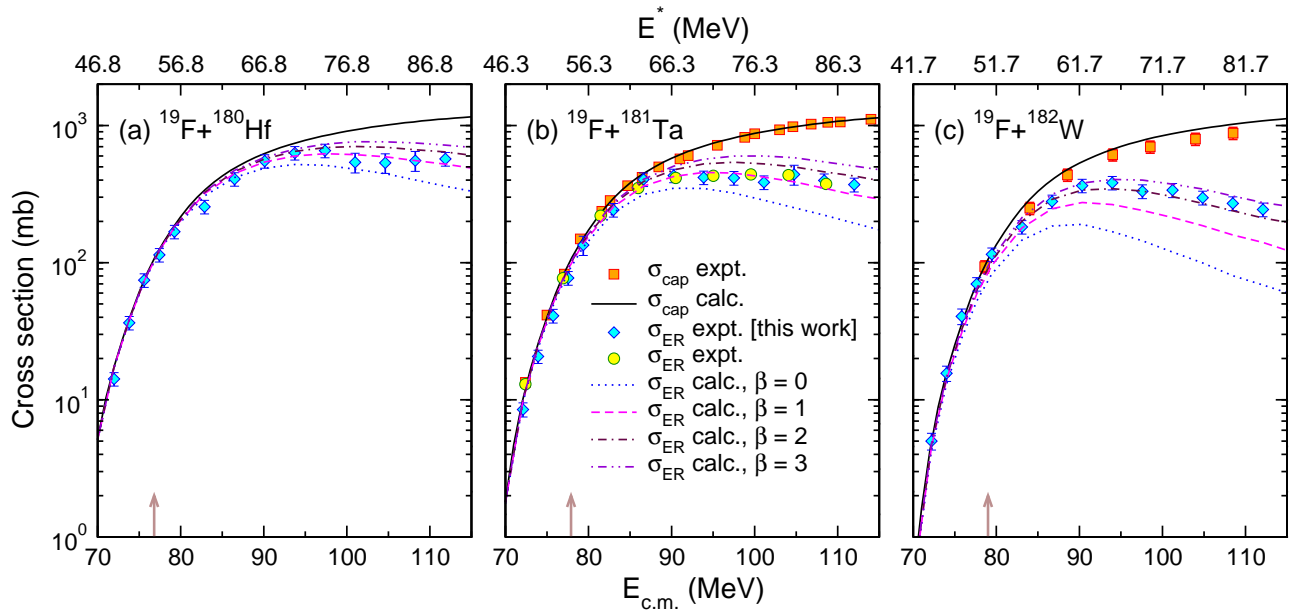}
\caption{\label{CrossSecs} Experimental and calculated $\sigma_{\textrm{ER}}$
for (a) $^{19}$F+$^{180}$Hf, (b) $^{19}$F+$^{181}$Ta and (c) $^{19}$F+$^{182}$W.
Theoretical capture cross sections, calculated by \textsc{ccfull}, are also
shown for each system. Data points represented by filled (yellow) circles are
obtained from Ref. \cite{RJCharity1986}. The vertical arrow in each panel
indicates the respective $V_{\textrm{B}}$.}
\end{figure*}

Measured $\sigma_{\textrm{ER}}$ for the three reactions are shown in Fig.
\ref{CrossSecs}. ER excitation function for $^{19}$F+$^{181}$Ta had been
reported earlier \cite{RJCharity1986}. Nevertheless we have measured
$\sigma_{\textrm{ER}}$ for this reaction, along with the same for the other two
reactions, to ensure similar systematic errors, if any, in measured data. Our
results for $^{19}$F+$^{181}$Ta are in agreement with the same reported in
Ref. \cite{RJCharity1986} within experimental uncertainties.

\subsection{Statistical model calculation}
The fate of a CN is decided in the present statistical model (SM) by following
its time evolution through Monte Carlo sampling of the decay widths of various
channels. Emission of neutrons, protons, $\alpha$-particles and $\gamma$-rays
along with fission are considered as the probable channels of decay. A CN can
undergo either fission with or without preceding evaporated particles and
photons or reduce to an ER. The final values of various observables are obtained
as averages over a large ensemble of events. The fission width is obtained from
the transition-state model of fission due to Bohr and Wheeler \cite{Bohr1939}
with certain modifications as outlined below. The particle and $\gamma$-decay
widths are obtained from the Weisskopf formula as given in Ref.
\cite{Frobrich1998}.  

We obtain the fission barrier in the present calculation by including shell
correction in the liquid-drop nuclear mass. Since the shell correction term
$\delta$ is defined as the difference between the experimental and the
liquid-drop model (LDM) masses ($\delta = M_{\textrm{exp}} - M_{\textrm{LDM}}$),
the full fission barrier $B_{\textrm{f}}(\ell)$ of a nucleus carrying angular
momentum $\ell$ is given as
\begin{equation}
B_{\textrm{f}}(\ell) = B_{\textrm{f}}^{\textrm{LDM}}(\ell) -
(\delta_{\textrm{g}} - \delta_{\textrm{s}})
\end{equation} 
where $B_{\textrm{f}}^{\textrm{LDM}}(\ell)$ is the finite-range liquid drop
model (FRLDM) fission barrier \cite{Sierk1986} and $\delta_{\textrm{g}}$ and
$\delta_{\textrm{s}}$ are the shell correction energies at the ground state
and the saddle configurations, respectively. The shell corrections at ground
state and saddle are obtained following the recipe given in Ref.
\cite{Myers1966} for including deformation dependence in shell correction
energy. 

It is usually assumed that the orientation of the CN angular momentum remains
perpendicular to both the reaction plane and the symmetry axis throughout the
course of the reaction and the LDM fission barrier thus is obtained for $K=0$,
where $K$ is the angular momentum component along the symmetry axis. However,
the initial CN angular momentum direction can change its orientation due to
perturbation by intrinsic nuclear motion \cite{Lestone2009}. Therefore, fission
barriers for $K\neq0$, which are larger than the $K=0$ barrier, are also to be
considered. This results in a reduction of the fission width which we have taken
into account following Ref. \cite{Lestone1999}.

The influence of shell structure in nuclear single-particle levels in the
nuclear level density which is used to calculate various decay widths of the CN
is obtained from the works of Ignatyuk et al. \cite{Ignatyuk1975} where the
following form of the level density parameter $a$ is given
\begin{equation}
a(E^{*}) = \tilde{a} \left[ 1 + \frac{g(E^{*})}{E^{*}}\delta \right]
\end{equation}
\noindent
where
\begin{equation}
g(E^{*}) = 1 - \exp \left( -\frac{E^{*}}{E_{\textrm{D}}} \right)
\end{equation}
\noindent
and $E_{\textrm{D}}$ is a parameter which determines the rate at which the shell
effect decreases with increase of $E^{*}$. The level density parameter is shape
dependent and its asymptotic form $\tilde{a}$ at high $E^{*}$ is taken from Ref.
\cite{Reisdorf1981}.

We next consider the collective enhancement of level density (CELD) which arises
due to the residual interaction giving rise to correlation among particle-hole
states resulting in collective excitations. The total level density
$\rho(E^{*})$ then can be written as \cite{BBM1974}
\begin{equation}
\rho (E^{*}) = K_{\textrm{coll}}(E^{*})\rho_{\textrm{intr}} (E^{*})
\end{equation} 
\noindent
where $\rho_{\textrm{intr}}(E^{*})$ is the intrinsic level density and
$K_{\textrm{coll}}$ is the collective enhancement factor. 

The rotational ($K_{\textrm{rot}}$) and vibrational ($K_{\textrm{vib}}$)
enhancement factors are taken from the work of Ignatyuk \textit{et al.}
\cite{Ignatyuk1985}. A smooth transition from $K_{\textrm{vib}}$ to
$K_{\textrm{rot}}$ with increasing quadrupole deformation $|\beta_{2}|$ of the
CN is obtained using a function $\varphi(|\beta_{2}|)$ given as follows
\cite{Zagrebaev2001}
\begin{subequations}
\begin{equation}
\label{kcol}
K_{\textrm{coll}}(E^{*}) = [K_{\textrm{rot}}\varphi(|\beta_{2}|)
+K_{\textrm{vib}}(1-\varphi(|\beta_{2}|))]\ f(E^{*}),
\end{equation}
\noindent
where
\begin{equation}
\varphi(|\beta_{2}|) = \left[1+\exp\left(\frac{\beta_{2}^{0}-|\beta_{2}|}
{\Delta\beta_{2}}\right)\right]^{-1}.
\end{equation}
\end{subequations}

The values $\beta_{2}^{0}= 0.15$ and $\Delta\beta_{2}= 0.04$ are taken from Ref.
\cite{Ohta2001}. The following form of the function $f(E^{*})$ accounts for the
damping of collective effects with increasing excitation \cite{ARJunghans1998}
\begin{equation}
\label{f_E}
f(E^{\textrm{*}})= \left[1+\exp\left(\frac{E^{*}-E_{\textrm{cr}}}
{\Delta E}\right)\right]^{-1}.
\end{equation}

The values of $E_{\textrm{cr}}$ and $\Delta E$ are taken as 40 MeV and 10 MeV,
respectively, which were obtained by fitting yields from projectile
fragmentation experiments \cite{ARJunghans1998}. The lowest value of
$K_{\textrm{coll}}(E^{*})$ is pegged at 1.

It is observed in numerous studies that a fission hindrance with respect to the
Bohr-Wheeler fission width is required in order to reproduce pre-scission
neutron multiplicity data from fusion-fission reactions (see \textit{e.g.}
Ref. \cite{Lestone2009}). A reduction in fission width is obtained from the
dissipative stochastic dynamical model of fission due to Kramers where the
fission width is given as \cite{Kramers1940},
\begin{multline}
\label{Kram}
\Gamma_{\textrm{f}}^{\textrm{Kram}} (E^{*}, \ell, K) \\
= \Gamma_{\textrm{f}}^{\textrm{BW}} (E^{*}, \ell, K)
\left\{ \sqrt{1+\left(\frac{\beta}{2\omega_{s}}\right)^{2}}-\frac{\beta}{2\omega_{s}} \right\},
\end{multline}
\noindent 
where $\beta$ is the reduced dissipation coefficient (ratio of the dissipation
coefficient to inertia) and $\omega_{\textrm{s}}$ is the frequency of a harmonic
oscillator potential which approximates nuclear potential in the saddle region.
In a stochastic dynamical model of fission, the fission rate reaches its
stationary value as given by Eq. \ref{Kram} after elapse of a certain time
interval \cite{Grange1983}. We therefore use a parametrized form of
time-dependent fission width as given in Ref. \cite{BhattPRC33}.

The above features are incorporated in a statistical model code \textsc{vecstat}
\cite{TBanerjee2018}. Detailed application of the model is discussed elsewhere
\cite{TBanerjee2019}.

Decay widths and fission barrier depend upon the angular momentum of the CN.
The $\ell$-distribution for capture are fed into the statistical
model as input. Total and partial capture cross sections
($\sigma_{\textrm{cap}}$ and $\sigma_{\ell}$, respectively) at a given energy
of the projectile can be calculated by coupled-channels formalism. To this end,
$\sigma_{\textrm{cap}}$ for $^{19}$F+$^{181}$Ta \cite{AKNasirov2010} and
$^{19}$F+$^{182}$W have been reproduced by the coupled-channels code
\textsc{ccfull} \cite{KHagino1999} incorporating appropriate potential
parameters and couplings. $\sigma_{\textrm{cap}}$ data for the latter reaction
have been obtained by adding $\sigma_{\textrm{ER}}$ (this work) and
$\sigma_{\textrm{fiss}}$ \cite{TBanerjee2017}. $\sigma_{\textrm{cap}}$ and
$\sigma_{\ell}$ for $^{19}$F+$^{180}$Hf have been calculated assuming potential
parameters and coupling scheme similar to the other two reactions.

In the present work, SM calculations are performed treating $\beta$ as the only
adjustable parameter. Results from SM calculation with different values of
$\beta$, along with results of coupled-channels calculation, are shown in Fig.
\ref{CrossSecs}.

\section{Discussion}
\label{Discuss}

\begin{figure*}[ht!]
\includegraphics[width=0.95\textwidth,trim=0.0cm 0.0cm 0.0cm 0.0cm,clip]{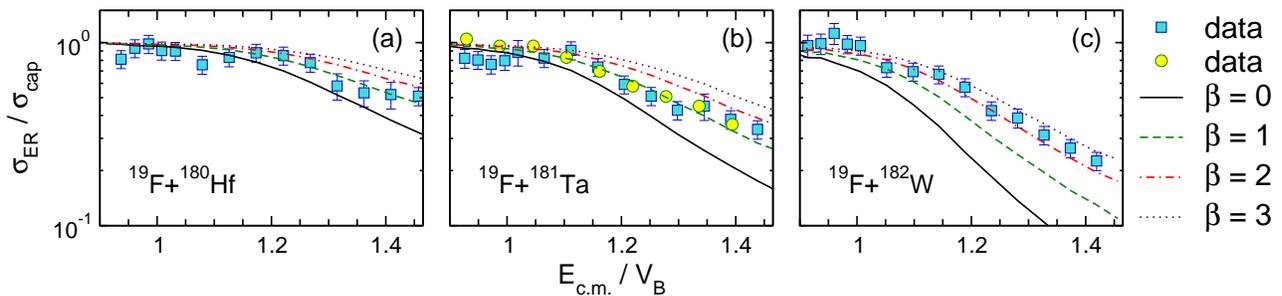}
\caption{\label{XSecRatio} Measured and calculated $\sigma_{\textrm{ER}}$
normalized by theoretical $\sigma_{\textrm{cap}}$ for (a) $^{19}$F+$^{180}$Hf,
(b) $^{19}$F+$^{181}$Ta and (c) $^{19}$F+$^{182}$W. Data points represented by
filled (yellow) circles are obtained from Ref. \cite{RJCharity1986}.}
\end{figure*}

Effects of shell closure on nuclear reaction dynamics have been investigated
through various observables. Enhanced FF anisotropy with respect to those
predicted by the standard statistical saddle point model, observed in
$^{12}$C+$^{198}$Pt \cite{AShrivastava1999}, was attributed to the effects of
the $N=126$ shell in the potential energy surface (PES) of the CN. On the other
hand, no signature of the modification of the PES due to the effect of $N=126$
shell closure was manifest in FF mass distribution \cite{AChaudhuri2015}, as
normalized width of mass distributions from $^{12}$C+$^{194,198}$Pt (leading to
CN with $N=120$ and 126, respectively) was found to be almost identical.
Influence of shell closure in the reaction partners on the mass and angle
distribution of FFs \cite{CSimenel2012}) and signatures of the $Z=82$ shell
closure in $\alpha$-decay process in heavy nuclei \cite{ANAndreyev2013} have
been reported in recent years.

Unlike FFs and neutrons, which may originate from both equilibrated CN or
non-equilibrium processes, ERs are the most unambiguous signatures of CN
formation. However, theoretical reproduction of $\sigma_{\textrm{ER}}$ is not
always free from uncertainties. For heavy fissile systems,
$\sigma_{\textrm{ER}}$ can be expressed as
\begin{equation}
\label{ERXSec}
\sigma_{\textrm{ER}}(E_{\textrm{c.m.}}) = \sum_{\ell=0}^{\infty}
\sigma_{\textrm{cap}}(E_{\textrm{c.m.}}, \ell)
P_{\textrm{CN}}(E_{\textrm{c.m.}}, \ell)
P_{\textrm{sur}}(E^{\textrm{*}}, \ell)
\end{equation}
where the three terms on the right hand side of Eq. \ref{ERXSec} denote (a)
probability of the collision partners to overcome the potential barrier in the
entrance channel, (b) probability that the composite system will evolve into an
equilibrated mononucleus starting from the touching configuration inside the
fission saddle point and (c) probability that the CN will survive as a cold ER,
respectively.

The second term on the right hand side of Eq. \ref{ERXSec}, is the least
precisely known. Considerable variance is also known to exist among the
different statistical models, which are frequently used to calculate the third
term on the right hand side of Eq. \ref{ERXSec}. Given these difficulties,
comparing the ER excitation functions of three similar reactions and looking
for signatures of $Z=82$ shell closure are quite challenging.

While trying to reproduce $\sigma_{\textrm{ER}}$ with the statistical model for
decay of CN, it is implicitly assumed that $P_{\textrm{CN}} = 1$. In other
words, the target-projectile composite system is assumed to yield an
equilibrated CN and not to proceed towards non-equilibrium fission-like
processes. This assumption is questionable. Several studies on presence of
non-equilibrium processes in 200 amu mass region have been reported. Shidling
\textit{et al.} interpreted reduction of $\sigma_{\textrm{ER}}$ in
$^{19}$F+$^{181}$Ta, compared to the same in $^{16}$O+$^{184}$W, as a
consequence of pre-equilibrium fission \cite{PDShidling2008}. Nasirov
\textit{et al.} \cite{AKNasirov2010} performed detailed analysis of these two
reactions within the framework of the di-nuclear system (DNS) model. According
to the results from the DNS model, quasi-fission and fast fission cause
hindrance to complete fusion in both reactions, albeit with varying degree of
severity. On the other hand, study of FF mass distribution did not find any
signature of quasifission for the reactions $^{19}$F+$^{181}$Ta and
$^{16}$O+$^{184}$W \cite{AChaudhuri2016}.

In the light of these conflicting reports, we argue that (a) presence of NCNF
in the three $^{19}$F-induced reactions under consideration is not significant
and (b) influence of NCNF, if any, on $\sigma_{\textrm{ER}}$ in these reactions
are comparable as the entrance channel parameters
$Z_{\textrm{p}}Z_{\textrm{t}}$, $\eta$ and structural features of the targets
are rather similar.

In reproducing observables from fusion-fission reactions, the input parameters
in the SM such as level density, fission barrier and fission delay time are
often varied in an \textit{ad hoc} manner. In the present work, no parameter of
the SM except for $\beta$ is varied to interpret the data. Fig. \ref{CrossSecs}
shows that while $\beta =$ 1\textendash2 $\times 10^{21}$ s$^{-1}$ reproduces
the ER excitation functions of $^{19}$F+$^{180}$Hf and $^{19}$F+$^{181}$Ta
systems over the entire range of excitation energy, higher values of $\beta =$
2\textendash3 $\times 10^{21}$ s$^{-1}$ are required for the $^{19}$F+$^{182}$W
system. Similar observations are also made in Fig. \ref{XSecRatio} where
measured and calculated $\sigma_{\textrm{ER}}$, normalized by
$\sigma_{\textrm{cap}}$ (obtained from coupled-channels calculations), are
plotted. The necessity of a higher value for $\beta$ for the $^{19}$F+$^{182}$W
reaction possibly arises from the facts that (a) the excitation energy of the CN
for this system is about 5 MeV less than those of the other two systems and (b)
the parameters deciding the energy dependence of CELD (Eq. \ref{f_E}) are not
optimized for the present systems but are taken from  an earlier work
\cite{ARJunghans1998}. The latter aspect requires further investigation in
future studies. However, the above values of $\beta$ are in agreement with the
theoretical estimate of pre-saddle dissipation strength based on the
chaos-weighted wall formula \cite{GChaudhuri2002}. It can also be noted from
Fig. \ref{XSecRatio} that $\frac{\sigma_{\textrm{ER}}}{\sigma_{\textrm{cap}}}$
reduces gradually with increasing $\chi_{\textrm{CN}}$. This is as expected
since fission becomes a more dominant decay mode in CN with larger fissility.

\section{Summary and Conclusion}
\label{Conc}
ER excitation functions have been measured for three reactions in similar
range of excitation energies in order to look for stabilizing effects of $Z=82$
shell closure against fission. The systems have been chosen in such a way that
the three CN, formed in these reactions, have same number of neutrons
($N = 118$) but different numbes of protons ($Z = 81$, 82, 83). A not-so-heavy
projectile ($A_{\textrm{p}} < 20$) has been chosen to ensure that the effect of
NCNF on ER formation is not severe. The three targets also have quite similar
structural features. Entrance channel parameters for the three reactions being
comparable, presence of NCNF, if any, is thus expected to affect ER formation
in the three reactions quite similarly. Measured cross sections have been
compared with statistical model predictions. The model includes shell effect in
level density, shell correction in fission barrier, $K$-orientation and CELD.
Reduced dissipation coefficient is the only adustable parameter. It is found
that the ER excitation functions can be reasonably reproduced with values of
$\beta$ in the range of 1\textendash3 $\times$ $10^{21}$ s$^{-1}$. The ratio
$\frac{\sigma_{\textrm{ER}}}{\sigma_{\textrm{cap}}}$
decreases with increasing fissility of the CN in the similar range of
excitation energies. No significant and abrupt deviations have been found in
the results obtained from $^{19}$F+$^{181}$Ta as an evidence in favour of
stabilizing effects of $Z = 82$ shell closure against fission. For further
validation of this conclusion, a more exclusive measurement of individual exit
channel cross sections in such reactions can be carried out in future.

\begin{acknowledgments}
The authors thank the Pelletron staff of IUAC for excellent support throughout
the experiment, Abhilash S. R. for assistance in fabricating
isotopically-enriched thin targets and Dr. E. Prasad for providing $^{180}$Hf
target.
\end{acknowledgments}

\end{document}